# A versatile apparatus for simultaneous trapping of multiple species of ultracold atoms and ions to enable studies of low energy collisions and cold chemistry

Bubai Rahaman[1], Satyabrata Baidya[1] and Sourav Dutta[1,*]

[1]*Tata Institute of Fundamental Research, 1 Homi Bhabha Road, Colaba, Mumbai 400005, India*
*Author to whom correspondence should be addressed: sourav.dutta@tifr.res.in

We describe an apparatus where many species of ultracold atoms can be simultaneously trapped and overlapped with many species of ions in a Paul trap. Several design innovations are made to increase the versatility of the apparatus while keeping the size and cost reasonable. We demonstrate the operation of a 3-dimensional (3D) magneto-optical trap (MOT) of $^7$Li using a single external cavity diode laser. The $^7$Li MOT is loaded from an atomic beam, with atoms slowed using a Zeeman slower designed to work simultaneously for Li and Sr. The operation of a 3D MOT of $^{133}$Cs, loaded from a 2D MOT, is demonstrated and provisions for MOTs of Rb and K in the same vacuum manifold exist. We demonstrate the trapping of $^7$Li$^+$ and $^{133}$Cs$^+$ at different setting of the Paul trap and their detection using an integrated time-of-flight mass spectrometer. We present results on low energy neutral-neutral collisions ($^{133}$Cs-$^{133}$Cs, $^7$Li-$^7$Li and $^{133}$Cs-$^7$Li collisions) and charge-neutral collisions ($^{133}$Cs$^+$-$^{133}$Cs and $^7$Li$^+$-$^7$Li collisions). We show evidence of sympathetic cooling of $^7$Li$^+$ ($^{133}$Cs$^+$) due to collisions with the ultracold $^7$Li ($^{133}$Cs).

## I. INTRODUCTION

Over the last three decades, the fields of cold trapped neutral atoms and trapped ions have individually seen tremendous progress enabling the precise control of the quantum state of these systems. This has intensified the research efforts in precision spectroscopy, quantum simulation, quantum computation and quantum logic spectroscopy [1–7]. Alongside, new directions have evolved which aim to achieve similar level of control in dipolar molecules [8–19] and in hybrid systems consisting of ions and atoms or ions and molecules [20–36]. The trend of studying increasingly complex systems perhaps originates from the "bottom-up" approach to understanding small systems and emergent phenomena.

While experiments with hybrid systems as a whole continue to produce trailblazing science, the apparatuses have become more sophisticated, specialized and goal-driven. This approach of course has the advantage of attaining superior performance for a specified scientific objective, but often such an apparatus leaves little room for exploratory and quickly adaptable inquisitive experimental research. As an example, once a research group chooses Cs as the atomic system and builds an apparatus, it is typically very difficult to switch to a different species, say Li, or to address some question that would require one to detect Cs$_2$ or Cs$^+$. Switching between species can be difficult because of restricted availability of laser sources or the lack of versatility in the vacuum apparatus. In this article, we focus on the latter and report a versatile apparatus that can be used to simultaneously trap and cool several species. This would allow one to quickly switch between species and open up the possibility of combining different species to explore new directions in cold chemistry. As one would expect, such an apparatus has limitations arising from the compromises one needs to make to accommodate different species with widely different properties. It is nevertheless worthwhile to invest in such apparatuses, especially in space-constrained and resource-constrained settings. In our laboratory, some of the planned research directions are: to study the formation of ultracold polar diatomic molecules, use them for studies of the $\pm C_3/r^3$ dipole-dipole interaction and emulate strongly interacting quantum systems [37,38]; detect the molecules by resonance enhanced multi-photon ionization (REMPI); detect, using REMPI assisted mass spectrometry, the long-lived molecular complexes conjectured to be formed in ultracold molecule-molecule collisions [39]; study collisions between trapped ions and atoms interacting via the $-C_4/r^4$ potential and explore cold charge exchange reactions [20]; study collisions between trapped ions and dipolar molecules interacting via the $-C_2/r^2$ potential and explore the effects of competing dipole-dipole and charge-dipole interactions.

Our goal is to build a single apparatus that can be used to address these research directions. In addition to increasing the range of possible questions that can be addressed, a single apparatus allows substantial monitory savings by reducing the number of vacuum pumps, vacuum gauges etc. and economical use of the laboratory space. We do not attempt to achieve the state-of-the-art performance for any of the individual species (e.g. in terms of number of atoms trapped, atomic density or atom loading rate) but rather focus on the versatility, while achieving reasonable performance for all species and maintaining ultra-high vacuum (UHV) conditions with pressure ~10$^{-11}$ mbar.

We report an apparatus which is designed to trap and laser cool neutral atoms of Li, K, Rb, Cs and Sr as well as trap ions in a Paul trap. We demonstrate the simultaneous and overlapped operation of 3-dimensional (3D) magneto-optical traps (MOTs) of $^7$Li and $^{133}$Cs. The 3D MOT of $^7$Li is loaded from an atomic beam, with atoms slowed using a Zeeman slower which is designed to work for Sr as well. The 3D MOT of $^{133}$Cs is loaded from a 2D MOT and the same vacuum manifold can be used for 2D MOTs of K and Rb. In the same region of space as the 3D MOTs, positively charged ions can be trapped in a four-rod linear

Paul trap. We demonstrate the trapping of $^7$Li$^+$ and $^{133}$Cs$^+$ by varying the electrical settings of the linear Paul trap. The electrodes of the Paul trap can be electrically configured in real time to exact the ions either along the axial or the radial direction and identify the ions using the in-built time-of-flight (ToF) mass spectrometers (MS). While we demonstrate the trapping of the lightest and the heaviest alkali-metal atoms and ions ($^7$Li, $^{133}$Cs, $^7$Li$^+$, $^{133}$Cs$^+$) only because of limited availability of laser sources, the vacuum apparatus can be used to trap K, Rb and Sr and their ions without the need to open the vacuum chamber. We present details of the apparatus in section II.

In section III, we demonstrate the applicability of the apparatus by presenting evidence of low energy collisions between trapped $^7$Li and $^{133}$Cs atoms and determining the atomic loss rates due to such collisions. We further present results on collisions between laser-cooled $^{133}$Cs ($^7$Li) and trapped $^{133}$Cs$^+$ ($^7$Li$^+$). We report signatures of sympathetic cooling of $^7$Li$^+$ by $^7$Li for the first time. We conclude with a discussion in section IV.

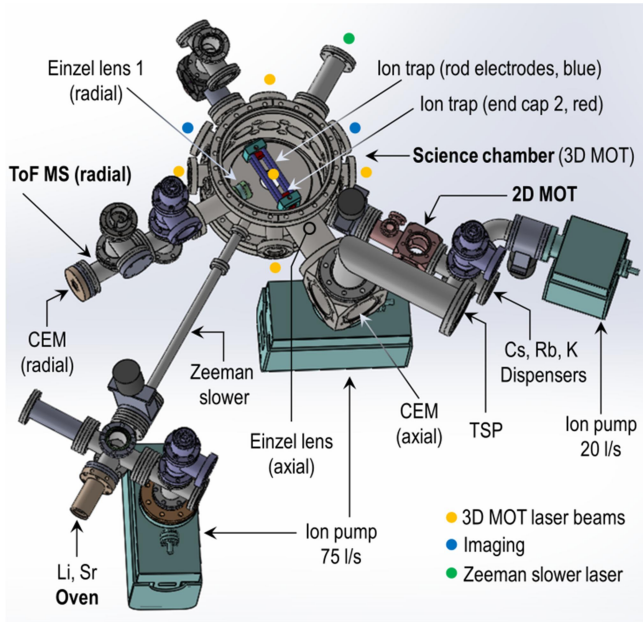

**FIG. 1.** Schematic overview of the vacuum apparatus is shown. Atoms and ions are trapped in the science chamber for studies of low energy collisions. The 2D MOT and the oven in combination with the Zeeman slower serve as sources of slowed atoms that are trapped in the 3D MOT. The orientations of the linear Paul trap, the axial ToF MS and the radial ToF MS are evident. For clarity, the top lid, Zeeman slower solenoid and the electro-magnetic coils (for the MOTs, for cancelling the Earth's field and for nulling the remnant field of the Zeeman slower) are not shown.

## II. DESCRIPTION OF THE APPARATUS

A sketch showing the overview of the vacuum apparatus is shown in Fig. 1. It consists of three building blocks: (*a*) a central "science chamber" that hosts the 3D MOTs and a linear Paul trap, (*b*) a 2D MOT section for generating slowed atomic beams for K, Rb and Cs, and (*c*) an oven section with a Zeeman slower for Li and Sr. The science chamber is constructed using non-magnetic stainless steel (SS316L). It consists of two CF200 ports oriented vertically, 12 ports (11 CF35 + 1 CF63) oriented horizontally and 16 CF16 ports inclined at ±11° with respect to the horizontal plane. The top and bottom lids consist of CF200-CF35 recessed adaptors on which CF35 optical viewports are mounted. The arrangement allows the electro-magnetic coils for the 3D MOT to be placed around the CF35 viewports such that the vertical separation between the coils is only 50 mm. The science chamber is connected to the 2D MOT section through a differential pumping tube (length 60 mm, inner diameter 2.5 mm) and to the oven section though the Zeeman slower tube (length 450 mm, inner diameter 16 mm). Inside the Zeeman slower tube, at the end closer to the oven, another tube of inner diameter 7 mm and length 90 mm is inserted to further reduce the vacuum conduction. The differential pumping maintains a pressure of $10^{-11}$ mbar in the science chamber even when the 2D MOT section and the oven are at $10^{-8}$ mbar pressure. The gate valves placed between the science chamber and the two other sections can be closed to separate the chambers if needed while refilling the atomic sources. A linear Paul trap is placed at the center of the science chamber and the trapped ions can be extracted either axially or radially for detection using the axial or radial ToF MS, respectively. In the rest of this section, we describe the various components of the apparatus and discuss the rationale behind the choices made. We indicate possible alternatives where ever possible to allow customization of the apparatus according to the need of the user.

### A. Atomic source

One of our primary considerations while deciding on the atomic sources was to maintain the pressure in the 3D MOT region at or below $10^{-11}$ mbar. There are two choices for atomic source: a dispenser source or an oven. We use commercially available dispenser sources (Alfavakuo, S-type, Bi-alloy) for K, Rb and Cs because they are readily available and easy to operate. The dispensers are mounted on a CF35 8-pin electrical feedthrough such that the orifices of the dispensers have direct line-of-sight with the 2D MOT region. This direct-line-sight allows the 2D MOT to be loaded directly from the effusive atomic beam. The dispenser orifices are located at a distance of ~10 cm from the center of 2D MOT. Closer placement of the orifices would be beneficial (because of higher atomic flux) but other geometric constraints restricted their placement. Too close a placement should, however, be avoided because the current through the dispenser creates a magnetic field that affects the operation of the 2D MOT. We typically operate the Cs dispenser at a current of 3 A, which keeps the pressure in the 2D MOT region below $1\times10^{-9}$ mbar. The K and Rb dispensers were activated during the vacuum bake-out by passing currents of ~ 3.5 A and 3.0 A, respectively, but they have otherwise not been operated in this work.

We use a dual-species home-built atomic oven for Li and Sr. The choice of oven is dictated by the unavailability of suitable dispenser sources and the high melting point, 180.5°C and 777°C for Li and Sr, respectively. Notably, the vapour pressures of Li and Sr at a given temperature are similar which is advantageous in designing a dual-species oven. However, it is still not advisable to mix two metals in an oven. To retain flexibility, we design a dual-species oven that can be adapted to other species with some straight-forward modifications. A sketch of the oven is shown in Fig. 2. The oven consists of two semi-cylindrical shaped chambers for the two species separated by a thin wall of thickness 0.5 mm. To construct the oven, a stainless steel (SS304) rod of diameter 70 mm and length 83 mm was machined to make a CF35 flange at one end and the diameter of the remaining rod (length ~70 mm) was reduced to 38 mm. Two semi-cylindrical cavities of diameter 14 mm, separated by a wall of 0.5 mm, were then machined using a wire-cut electric discharge machine. From the rear end (i.e. the end opposite to CF35 flange), material was then removed to increase the volume of the semi-cylindrical oven chambers (see Fig. 2) and two separate semi-cylindrical discs were welded to seal the rear end. On the front end containing the CF35 flange, the thin wall separating the two chambers was removed up to a depth of 10 mm by a milling machine. This was done to accommodate a nozzle for the atomic beam. Before inserting the nozzle, solid metallic pieces of Li and Sr were inserted into the left and right compartments of the oven under an inert Argon atmosphere.

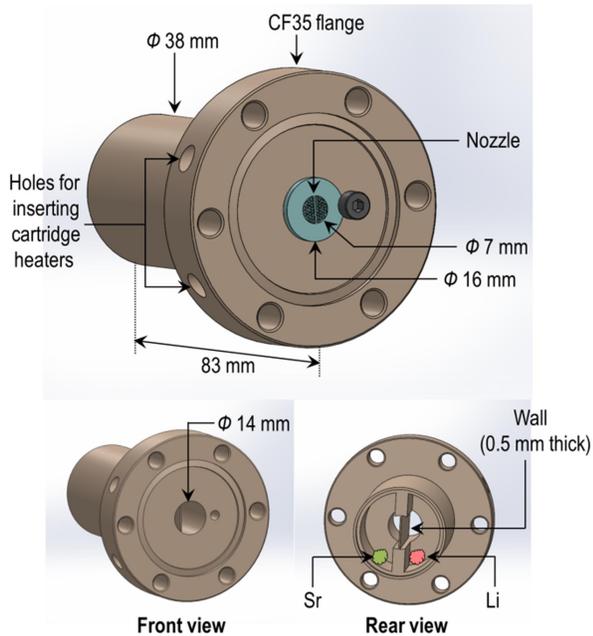

**FIG. 2.** Schematic diagram of the assembled dual species oven along with the nozzle containing hypodermic needles is shown in the top panel. The bottom panels show the front and rear views of the oven. In the rear view, the two welded semi-cylindrical discs that seal the oven are not shown for clarity.

The nozzle, shown in blue colour in Fig. 2, is made of SS304 and can slide to fit in the hole in the CF35 flange of the oven assembly and is clamped to the CF35 flange with a M4 screw. The nozzle has a total length of 12 mm and inner diameter (ID) of 7 mm separated into two parts by a 0.5 mm wall. The nozzle is packed with hypodermic needles of length 12 mm, OD 0.82 mm and ID 0.51 mm. The hypodermic needles are used to provide a collimated beam with low divergence (half angle ~ 2.4°). It is to be noted that the entire assembly of the oven and nozzle is made of SS304 and the copper gasket used to mount the CF35 flange of the oven on the vacuum chamber (i.e. the 6-way cross in Fig. 1) does not come in direct contact with the atomic beam. This reduces the changes of vacuum leaks arising from corrosion of copper. Another important design feature is that the nozzle + oven are heated using cartridge heaters inserted into the CF35 flange (see Fig. 2) to ensure that the nozzle always stays at a higher temperature compared to the oven, thus preventing the nozzle from clogging. In practice, we found it more convenient to use an additional heating tape wrapped around the oven to control the temperature of the oven while still keeping it below the nozzle temperature. We typically operate the oven at a temperature of ~310°C which is significantly lower than earlier reports where the oven is operated at ~500°C [40,41].

In our case, the two compartments of the oven containing Li and Sr are heated to the same temperature which is acceptable because their vapour pressures are similar. To use metals with different vapour pressures, the design of the oven should be modified so that the lengths of the two compartments are different i.e. longer compartment for the metal with lower vapour pressure. Taking into consideration the fact that the vapour pressure is determined by the coldest spot in the oven compartment, a standard thermal flow simulation package can be used to determine the optimum lengths.

The oven is connected to a 6-way cross which is pumped by an ion pump with a nominal pumping speed of 75 l/s. The pressure, determined from the current in the ion pump, is $\sim 2 \times 10^{-8}$ mbar. However, we believe that the ion pump is contaminated with Li and Sr vapour leading to higher ion current and higher pressure readings although the actual pressure is below $10^{-9}$ mbar. On one port of the 6-way cross, a manually rotatable atomic beam shutter is mounted using a rotary feedthrough. The shutter is manually rotated to allow or block the atomic beam from entering the Zeeman slower section.

**B. Zeeman slower**

At the operating temperature of the oven, the most probable speed of the Li and Sr atoms in the atomic beams are ~1170 m/s and ~330 m/s, respectively. The speed of the atoms needs to be reduced so that they can be captured in the 3D MOT. This is accomplished using a Zeeman slower in decreasing-field configuration [42]. The Zeeman slower solenoid consists of 9 sections, 8 of which are of length 35 mm each and 1 has a length of 12 mm. Enamelled copper wires of nominal diameter 1.63 mm (SWG16) were used to construct the solenoid. The same current flows through each section. The 9-section bobbin on which the wires are wound is made of aluminium and has provisions for cooling by flowing chilled water along the inner surface of the bobbin.

The optimal number of turns of wire in each section was determined using computer simulations. In the simulation results for Li shown in Fig. 3, we fixed the 671-nm laser detuning ($\Delta_{Li}$) at -218 MHz from the $2s\ ^2S_{1/2} \to 2p\ ^2P_{3/2}$ transition, the laser intensity saturation parameter ($I/I_{sat}$) at 6 and the current at 7 A. Here $I_{sat}$ is the saturation intensity for circularly polarized light. The simulations suggest that Li atoms at speed up to ~ 830 m/s can be slowed to ~ 250 m/s, assuming a fixed effective acceleration of $0.74 a_{max}$. Here, $a_{max} = \Gamma(\hbar k/2m)$ is the maximum possible deceleration of the atom when the laser is exactly on resonance with the $2s\ ^2S_{1/2} \to 2p\ ^2P_{3/2}$ transition, over the entire length of the Zeeman slower. In practice the final slowed atomic speed could be higher or lower depending on the effective deceleration experienced by the atoms. Figure 3 also shows that the measured magnetic field agrees very well (i.e. within a few gauss over a length of 30 cm) with the ideal theoretical magnetic field and the simulated magnetic field. Our simulations show that the same magnetic field configuration can slow Sr atoms from 520 m/s to 65 m/s, with a laser detuning of -110 MHz from the $5s^2\ ^1S_0 \to 5s5p\ ^1P_1$ transition and laser intensity saturation parameter $I/I_{sat} = 3$, assuming an effecting deceleration of $0.48 a_{max}$.

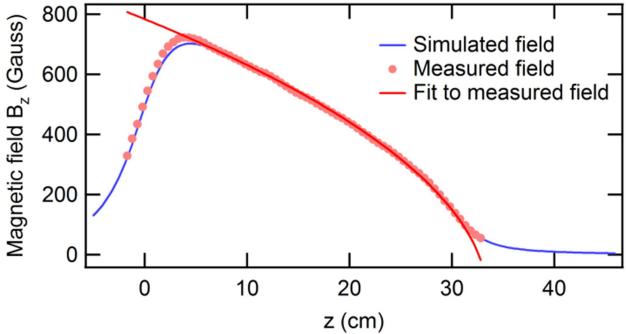

**FIG. 3.** The simulated, measured and fitted magnetic field ($B_z$) along the axis of the Zeeman slower is shown. The winding of the electro-magnet starts at z = 0. The MOT is located at z = 47 cm. The remnant field (~ 2 Gauss) of the Zeeman slower at the location of the MOT is cancelled using a separate shimming coil in the experiment.

In our regular day-to-day operation, we operate the Zeeman slower at a current of 3.5 A that generates a magnetic field of ~ 350 Gauss at the beginning of the slower. Simulations show that atoms moving with speed up to 490 m/s can be slowed. At this setting the heat dissipated is low and water cooling of the Zeeman slower is not required. The $^7$Li Zeeman slower laser beam has a total power of ~ 10 mW i.e. combined power of cooling and repumping beams (see discussion later) and is weakly focused to match the profile of the slightly diverging atomic beam. With these very modest laser power and magnetic field, we generate sufficient flux of slow $^7$Li atoms to load a 3D MOT with ~$10^8$ atoms.

**C. Laser systems**

We use a home-built external cavity diode laser (ECDL) for the $^7$Li MOT. The details of the ECDL have been reported elsewhere [43]. The ECDL delivers up to 150 mW of optical power and is frequency stabilized by locking to $2s\ ^2S_{1/2}\ (F=2) \to 2p\ ^2P_{3/2}\ (F')$ feature in the saturated absorption spectrum (SAS) of $^7$Li vapour in a heat-pipe-oven operated at ~315°C. Here, $F$ and $F'$ denote the hyperfine levels of the $s$ and $p$ states, respectively. The different $F'$ levels in the SAS of $^7$Li are not resolved. The detunings of the MOT and the Zeeman slower beams are controlled using acousto-optic modulators (AOMs) as follows. A small fraction of the laser output power (~ 10 mW) is picked off, sent through a 109-MHz AOM in double-pass configuration and used to obtain the SAS for laser frequency stabilization. The remaining light, detuned by -218 MHz from the $2s\ ^2S_{1/2}\ (F=2) \to 2p\ ^2P_{3/2}\ (F=3)$ transition, is first sent through an electro-optic modulator (EOM) operating at ~803 MHz to generate the frequency sideband to address the $2s\ ^2S_{1/2}\ (F=1) \to 2p\ ^2P_{3/2}\ (F')$ repumping transition. The cooling to repumping power ratio is approximately 2:1. The beam is then split into two parts. One part is coupled directly into an optical fiber and the output light (~10 mW) is used as the Zeeman slower laser beam. The other part is send through a 197-MHz AOM is single-pass configuration to generate the MOT light, detuned by -21.5 MHz from the $2p\ ^2P_{3/2}\ (F=3)$ level, and then coupled into an optical fiber to transport the MOT light close to the UHV chamber. Around 22 mW of power (cooling + repumping) is available for the $^7$Li MOT.

For the 2D and 3D MOTs of $^{133}$Cs, we use a commercial amplified laser system to address the $6s\ ^2S_{1/2}\ (F=4) \to 6p\ ^2P_{3/2}\ (F'=5)$ cooling transition and a homebuilt ECDL to address the $6s\ ^2S_{1/2}\ (F=3) \to 6p\ ^2P_{3/2}\ (F'=4)$ transition. SAS of $^{133}$Cs is used for frequency stabilization of the lasers. The repumping laser is locked to the $F=3 \to F'=4$ transition and the cooling laser beam locked slightly detuned from the $F=4 \to F'=5$ transition. To detune the cooling laser from the $F=4 \to F'=5$ resonance in a controllable manner, we use the Zeeman effect to our advantage. The vapour cell is housed inside a solenoid which is used to apply a controlled magnetic field to shift the atomic resonance and thereby control the detuning of the cooling laser. This method of locking and detuning the laser avoids the use of AOMs for frequency shifting and completely eliminates the optical power losses of AOMs (typically single-pass AOM efficiencies are ~ 80%). Additionally, the laser frequency can be tuned over a large range (~ 100 MHz) without changing the power and direction of the laser beam. In our case, however, since the same laser is used for both the 2D and 3D MOTs, we lose the capability to independently control the detunings. We found that a detuning of -10

MHz results in the maximum number of $^{133}$Cs atoms in the 3D MOT. The cooling and repumping laser beams are individually coupled into optical fibers and transported close to the UHV chamber where they are expanded and combined.

**D. 2D MOT section**

A 2D MOT [44–47] is used to generate slow atoms that can be trapped in the 3D MOT. A custom made modified CF35 6-way cube serves as the vacuum chamber for the 2D MOT. Out of the six CF35 ports, four ports have optical viewports (clear diameter 38 mm) for the 2D MOT laser beams, one port is connected to a 6-way cross (which hosts the connection to an ion pump, the dispenser sources discussed in section II.A., an angle valve and an optical viewport for a "push beam") and the remaining port is connected to the science chamber via the differential pumping tube and a gate valve. The differential pumping tube (length 60 mm, inner diameter 2.5 mm) is welded on a double-sided CF35 flange, which is sandwiched between the 6-way cube and the gate valve. The differential pumping tube limits the vacuum conductivity to the extent that the science chamber pressure does not change even when the 2D MOT section pressure reaches mid $10^{-9}$ mbar. In hindsight, a better alternative would have been a tube with a gradually increasing inner diameter which, as the cost of slightly increased vacuum conductivity, would prevent the loss of slowed atoms due to collisions with the inner walls of the tube.

We use laser beams of circular cross section with $1/e^2$–diameter ~ 22 mm for transverse cooling of $^{133}$Cs in the 2D MOT. The laser beams are slightly focussed to counteract the optical losses in the viewport so that the incident and retro-reflected beams have the same intensity inside the vacuum chamber. The peak intensity of each of the 4 laser beams for transverse cooling (repumping) is ~ 80 mW/cm$^2$ (~ 6 mW/cm$^2$). The 2D magnetic field gradient is generated by two pairs of square-shaped coils in anti-Helmholtz configuration. The magnetic field is zero over a length of ~ 15 mm along the axis of 2D MOT that coincides with the axis of the differential pumping tube and increases approximately linearly on moving away transversely from the axis. Each coil is wound on a square-shaped bobbin (inner side dimension 33 mm) and contains ~ 410 turns of insulated copper wire of diameter 1.1 mm. The separation between the coils in each pair is 115 mm. The magnetic field gradient generated is 6.4 (Gauss/cm)/A. The optimal magnetic field gradient for the 2D MOT is 14 Gauss/cm which we experimentally determined by maximizing the number of atoms in the 3D MOT. The 2D MOT is loaded directly from the effusive atomic dispenser source. We observed that a small 3D MOT can be loaded even in the absence of the magnetic field gradient as long as the transverse laser beams of the 2D MOT are kept operational. In future the 2D MOTs for K and Rb can be implemented by combing the respective laser beams with those for Cs using dichroic mirrors.

**E. 3D MOTs of $^7$Li and $^{133}$Cs**

To ensure overlap of the laser beams, the light for the $^7$Li MOT is combined with the cooling light for the $^{133}$Cs MOT using a dichroic mirror. The beams are then expanded in diameter and split into three parts using combinations of half-wave plate and polarization beam splitter (PBS). The $1/e^2$–diameters of the Cs and Li laser beams are 16 mm and 8.4 mm, respectively. The MOTs operate with 3 pairs of retro-reflected cooling laser beams. The single beam peak intensities of the Cs and Li cooling laser beams are 27 mW/cm$^2$ and 13.3 mW/cm$^2$, respectively. The required circular polarization is obtained using achromatic quarter-wave plates. The $^7$Li repumping laser beam, generated using an EOM, propagates with the $^7$Li cooling and is therefore applied from all six directions. The Cs repumping laser beam is applied along all six directions but two of them have circular polarization opposite to that of the cooling beams. This non ideal polarization of the Cs repumping beam is a compromise arising from difficulties in combining multiple laser beams, while ensuring the maximum available laser powers for the MOTs. We found that optimal magnetic gradient for the $^7$Li MOT is 16.8 Gauss/cm, while that for the $^{133}$Cs MOT is 11.2 Gauss/cm. To operate both MOTs simultaneously, we set the magnetic field gradient at 14.0 Gauss/cm. The magnetic field gradient is generated by two electro-magnets placed around the recessed CF35 viewports of the top and bottom lids of the science chamber. Each coil has an ID 70 mm and contains 112 turns of enamelled copper wire of diameter 1.1 mm. The vertical separation between the coils is 50 mm. The coils produce a magnetic field gradient of 5.6 (Gauss/cm)/A.

The MOT is monitored using fluorescence imaging. The fluorescence from the MOTs is collected with a lens and divided into two parts – one part is send to a CCD camera to record images from which the shape, size and location of the MOT is determined while the other part is sent to a silicon photodiode which provides an output current proportional to the number of atoms. Dichroic mirrors and optical filters are used to selectively record the images / photodiode signals of the $^7$Li and $^{133}$Cs MOTs. When operated individually (i.e. not overlapped), we typically can load ~ $1\times10^8$ (~ $1\times10^8$) atoms in the $^7$Li MOT ($^{133}$Cs MOT) when operating at the previously mentioned laser powers, detunings, magnetic field gradients, dispenser currents, oven temperatures etc. We have tested that the number of atoms increases by a factor of ~ 3 (~ 5) if the Li oven temperature (Cs dispenser current / 2D MOT pressure) is increased to 375°C (3.3 A / $4\times10^{-9}$ mbar).

**F. Ion trap**

The linear Paul trap (LPT) is positioned with its center coinciding with the center of the science chamber, thus coinciding with the center of the 3D MOTs. A sketch of the linear Paul trap is shown in Fig. 4. The linear Paul trap consists of four parallel titanium rods (diameter 8 mm, length 106 mm) and two ring-shaped SS316 end cap electrodes (ID 16 mm, OD 22 mm, length 30 mm). These are mounted on an electrically insulating holder made of machinable ceramic MACOR. The rods are placed in a square geometry such that the center-to-center distance between adjacent rods is 24 mm (see Fig. 4 inset). The 16 mm separation between the surfaces of adjacent rods allows the MOT laser beams to pass through. The distance between the end caps is 88 mm. The large separation between the various electrodes is beneficial in terms of optical access but has the disadvantage that the depth of the ion trap (< 1 eV) and the secular frequency (< 0.5 MHz) are low for practically applicable rf voltage. Radio-frequency (rf) voltages are applied to the rod electrodes such that diagonal electrodes have the same voltage while adjacent electrodes have voltages 180° out of phase. The rf is sourced from a function generator and amplified using a power amplifier (Krohn-hite 7602M, Frequency DC – 1 MHz, voltage up to ±200 V). Equal positive dc voltages are applied to the end cap electrodes. In our experiment, we found that optimal trapping of $^7Li^+$ ($^{133}Cs^+$) occurs at a rf frequency of 880 kHz (270 kHz), rf voltage amplitude of 60 V (140 V) and end cap voltage of 40 V (40V).

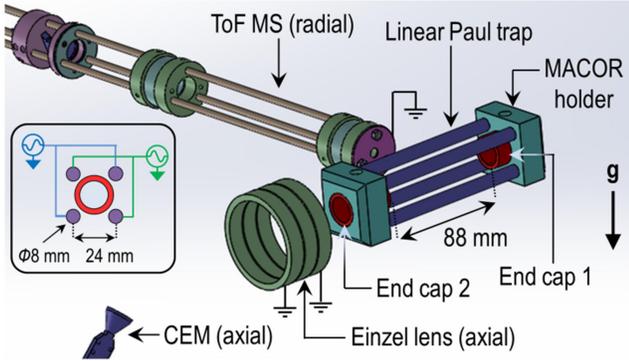

**FIG. 4.** Schematic diagram showing the linear Paul and the position of the Einzel lenses. The inset shows the cross-section view of the linear Paul trap and the rf voltage configuration for trapping ions.

We performed computer simulations of ion trajectories using SIMION to determine the range of rf frequencies and voltages required to trap ions of specific masses. At the experimental operating settings of $^7Li^+$ ($^{133}Cs^+$), the radial secular frequency is ~ 190 kHz (~ 90 kHz), axial secular frequency is ~ 38 kHz (~ 10 kHz) and the radial trap depth is ~ 0.2 eV (~0.5 eV), as determined from simulations. For the operating rf frequency of 880 kHz (270 kHz), we found that any rf voltage in the range 20 – 80 V (20 – 150 V) and any dc voltage up to +100 V can stably trap $^7Li^+$ ($^{133}Cs^+$). Through the simulations we also found that there are frequency/voltage settings that allow stable trapping of $^{39}K^+$, $^{87}Rb^+$ and $^{133}Cs^+$ simultaneously, but did not find any setting within the operating range of the rf amplifier where $^7Li^+$ can be trapped simultaneously with $^{133}Cs^+$. Given that we trap the lightest and the heaviest alkali-metal-ions in the experiment and the operating parameters are in agreement with simulation, we presume it should be possible to trap any species of intermediate mass in the ion trap by appropriately choosing the rf and DC voltages.

### G. Time-of-flight mass spectrometer

Singly charged ions of alkali-metal atoms in their ground state are optically inactive because the photon energy required to excite to the first electronically excited state is >13 eV, for which lasers (wavelength < 100-nm) are not available. Therefore, such ions cannot be detected using laser induced fluorescence. Molecular ions too are not amenable to fluorescence based detection. One way to detect and identify such ions is using time-of-flight (ToF) mass spectrometry. It works irrespective of whether or not the ions are trapped, although we will focus here on detection of trapped ions [48–52].

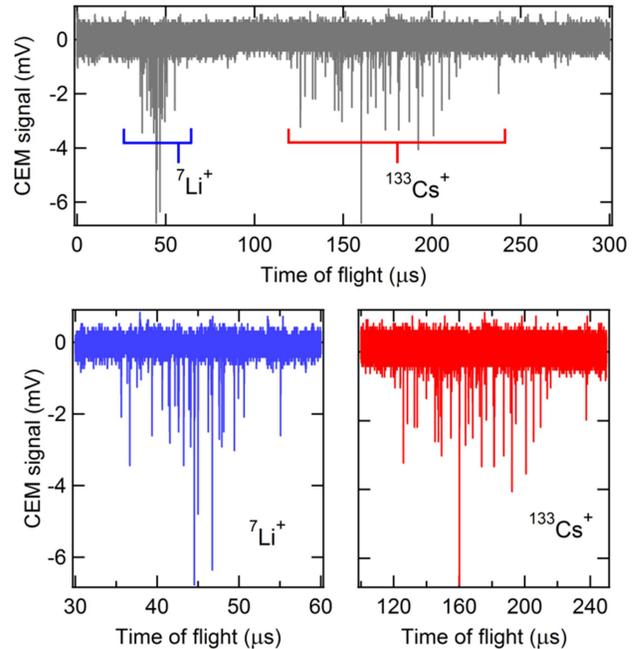

**FIG. 5.** The time of flight spectra of $^7Li^+$ and $^{133}Cs^+$ detected using the axial CEM is shown in the top panel. Zoomed in versions are shown in the lower panel. To record such spectra, the dc voltage on end cap #2 is suddenly switched to zero at t = 0. Each individual spike corresponds to a single ion hitting the CEM. For counting ions, the discriminator level is set at -1.5 mV. Note: the $^7Li^+$ and $^{133}Cs^+$ spectra were recorded at different values of rf and dc voltages because $^7Li^+$ and $^{133}Cs^+$ cannot be trapped simultaneously and then overlaid.

In our setup, trapped ions can be extracted from the LPT either axially (i.e. along the axis of the LPT) or radially (i.e. perpendicular to the axis of the LPT). The axial ToF-MS has high detection efficiency while the radial TOF-MS has high mass resolution. For axial extraction, the dc voltage on one of the end cap electrodes (end cap #2) is switched to zero within 100 ns, which accelerates the ions towards a channel electron multiplier (CEM) placed behind the end cap. The rf voltages are kept on. The axial CEM is located at a distance of 262 mm from the center of the science chamber and is displaced 10-mm vertically downwards to allow unhindered optical access to the center of the science chamber. The axial CEM has an extended dynamic range to allow for detection of larger number of ions without being saturated. No grounded metallic mesh or grid is placed in front of the CEM but the vacuum chamber is explicitly grounded. Through SIMION simulations we found that the grounded chamber plays a critical role in defining the electric field required to deflect the ions downwards so that they hit the CEM. In hindsight, an additional deflector plate to repel the ions towards the CEM might have been beneficial [53]. Simulations also showed that the mass resolution is poor although $^{7}$Li$^{+}$ and $^{133}$Cs$^{+}$ can be easily distinguished using their ToF. An Einzel lens, with three rings of inner diameter 52 mm and length 10 mm mutually separated by 1 mm, is placed between the end cap #2 and the axial CEM, 125 mm from the center of the science chamber, for focussing of ions. The central ring is kept at a dc potential of -56 V (-40 V) for optimal detection of $^{7}$Li$^{+}$ ($^{133}$Cs$^{+}$) and the two outer rings are grounded. We connect the CEM output directly to an oscilloscope where arrival of individual ions is detected in form of voltage pulses of duration ~10 ns as shown in Fig. 5. We record the signal on the oscilloscope and determine the number of ions by setting a discriminator level in the post-processing of the data.

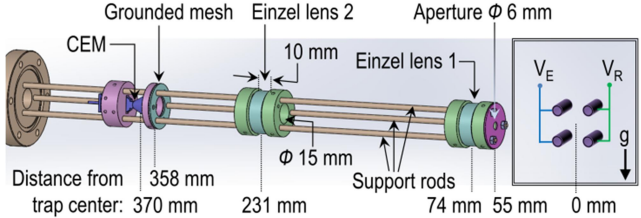

**FIG. 6.** Schematic diagram of the radial time-of-flight mass spectrometer along with the relative distances of the various components are shown. The boxed region shows the electrical configuration of the linear Paul trap electrodes during extraction of ions for detection. The central rings of the Einzel lenses are biased with positive voltages. The outer rings of the Einzel lenses, the aperture of diameter 6 mm and the support rods are electrically grounded.

The radial extraction of ions is more challenging because the rod electrodes need to serve as extraction electrodes as well. This necessitates turning off the rf voltages and then applying dc voltages to the rod electrodes such that the two rods closest to the radial Einzel lens #1 act as extraction electrodes (with dc voltage $V_E$) and the other two rods act as repeller electrodes (with dc voltage $V_R$). The radial ToF MS and the voltage configurations of the electrodes required extraction of ions are shown in Fig. 6. Our SIMION simulations predict $V_R \sim$ +1000 V and $V_E \sim$ +700 V for optimal detection of ions. Upon extraction, the ions diverge horizontally and vertically with different divergence angles. Two Einzel lenses, one located at 74 mm and the other at 231 mm from the center of the ion trap, are placed for focussing the ions and steering them to the CEM located at 370 mm. The voltage on each Einzel lens is ~ +300 V. A grounded mesh is placed in front of the CEM. We simulated the TOF of various ions by applying different voltages on the rod electrodes and the Einzel lenses and found that the mass resolution $m/\Delta m$ is > 100, which is sufficient to distinguish any species that may be formed our experiment.

For optimal extraction of the ions, it is necessary to switch off the rf voltages within a fraction of the rf time-period and then apply high voltage (HV) dc to the rod electrodes with a rise time below 100 ns. A complication arises because the rf voltage and HV dc are applied to the same electrodes and switching on/off generates transients that can damage the rf and dc sources. We are currently developing the electrical circuit to accomplish this.

### III. EXPERIMENTAL RESULTS

In this section we report on experimental results obtained using the apparatus. We discuss the measurement of low energy collisions between neutral $^{7}$Li and $^{133}$Cs atoms in a dual species MOT, the measurement of collisions between charged $^{7}$Li$^{+}$ ($^{133}$Cs$^{+}$) and neutral $^{7}$Li ($^{133}$Cs), and the measurements showing evidence of sympathetic cooling of $^{7}$Li$^{+}$ ($^{133}$Cs$^{+}$) by $^{7}$Li ($^{133}$Cs).

#### A. Collisions between neutral Li and Cs atoms

To study collisions between $^{7}$Li and $^{133}$Cs atoms at sub milli-Kelvin temperatures, we simultaneously laser-cool and trap $^{7}$Li and $^{133}$Cs atoms in their respective MOTs. The overlap of the two atomic clouds is ensured by carefully overlapping the MOT laser beams for the two species. Capabilities of small adjustments of the MOT position using shimming magnetic fields exist in the apparatus but were not needed. Collisions between atoms in a MOT lead to loss of the atoms if the kinetic energy post collision is higher than the effective trap depth of the MOT, which is typically ~ $k_B$×(few Kelvin) ~ $10^{-4}$ eV [54–56]. In a dual-species MOT of species $A$ and $B$, both homo-nuclear and hetero-nuclear collisions are prevalent and the respective loss rate coefficients, e.g. $\beta_A$ and $\beta_{A,B}$, can be determined. The loss rate coefficient ($\beta$) is related to the collision cross-section ($\sigma$) through the expression $\beta = \langle \sigma v \rangle$, where $v$ is relative collision speed. In a MOT, $v$ is typically of the order 1 m/s and therefore a MOT provides a good platform to measure low energy collision cross sections.

The collision rate coefficient $\beta_{A,B}$, describing the loss of atoms of species $A$ due to collisions with species $B$, can be determined by describing the loading dynamics of the MOT of species $A$ using a rate equation model [57,58]:

$$\frac{dN_A}{dt} = L_A - \gamma_A N_A - \beta_A \int n_A^2 \, d^3r - \beta_{A,B} \int n_A n_B \, d^3r \quad [1]$$

where $N_A$ is the number of atoms in the MOT of species $A$, $n_A$ and $n_B$ are the atomic number densities of species $A$ and $B$, respectively. $L_A$ is the loading rate of species $A$, $\gamma_A$ is the one-body loss rate of species $A$ due to collisions with background gases, $\beta_A$ is the loss rate coefficient that accounts for loss of species $A$ due to collisions with other atoms in the MOT of species $A$ and $\beta_{A,B}$ is the loss rate coefficient that accounts for loss of species $A$ due to collisions with atoms of species $B$. We are primarily interested in determining $\beta_{A,B}$ which is related to the collision cross-section $\sigma_{AB}$ through the expression $\beta_{A,B} = \langle \sigma_{AB} v_{AB} \rangle \approx \bar{\sigma}_{AB} \bar{v}_{AB}$, where $\bar{v}_{AB}$ is the mean relative speed between $A$ and $B$, and $\bar{\sigma}_{AB}$ is the collision cross-section at a collision speed of $\bar{v}_{AB}$. Two methods can be used to determine $\beta_{A,B}$: (*i*) recording the loading dynamics of species $A$ in presence of species $B$ or (*ii*) recording, in presence of species $B$, the decay in the number of atoms of species $A$ by setting $L_A$ to zero. We choose the latter because it is automatically immune to fluctuations in $L_A$ and results in a more robust analysis. In the experiment, we record the decay in the number of atoms of species $A$ ($^{133}$Cs in our case) either in absence or presence of the MOT of species $B$ ($^{7}$Li in our case) as shown in Fig. 7 and determine $\beta_{A,B}$ ($\beta_{Cs,Li}$ in our case) by fitting the data as discussed below.

The analysis using Eq. (1) is simplified if we make the following assumptions. (*i*) The MOT operates in the temperature-limited regime where the MOT volume $V_A$ remains fixed even when the number of atoms in the MOT changes. For a Gaussian atomic density distribution $n_A = n_{A0} \exp(-r^2/r_A^2)$ with $1/e$–radius $r_A$, we can write $\beta_A \int n_A^2 d^3r = \beta_A N_A^2/\sqrt{8} V_A$ where $V_A = (\pi r_A^2)^{3/2}$ and $n_A = N_A/V_A$. (*ii*) The MOT of species $A$ is much smaller in size and completely immersed in the MOT of species $B$ of density $n_{B0}$ that is constant i.e. $n_{B0}$ does not change with time and is independent of the presence or absence of the MOT of species $A$. This allows us to write $\beta_{A,B} \int n_A n_B d^3r \approx \beta_{A,B} n_{B0} \int n_A d^3r = \beta_{A,B} n_{B0} N_A$.

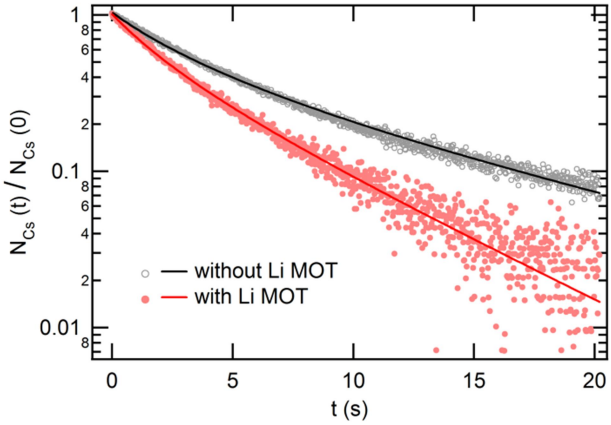

**FIG. 7.** The decay of the number of $^{133}$Cs atoms either in absence (open circles) or presence (filled circles) of the $^{7}$Li MOT. The solid lines are fits of the data to Eq. (3).

To check if assumption (*i*) is satisfied in the experiment, we recorded images of the $^{133}$Cs MOT while it is decaying and found the volume occupied by the atoms remain approximately constant i.e. the variation is less than 20%. To satisfy condition (*ii*), we ensure that the size of the $^{7}$Li atomic cloud [$1/e$–radius ~ 1.0 mm] is much larger than the $^{133}$Cs atomic cloud [$1/e$–radius ~ 0.44 mm]. To ensure that the density $n_{Li0}$ of the $^{7}$Li MOT remains approximately constant when the $^{133}$Cs MOT loads, we use a $^{7}$Li MOT with ~$10^8$ atoms, much larger than the number of $^{133}$Cs atoms (~$10^7$), so that the number of $^{7}$Li atoms lost due to binary $^{7}$Li-$^{133}$Cs collisions is small (~ 5%) compared to the total number of $^{7}$Li atoms. We note that the deviations in the experiment from idealistic assumptions (*i*) and (*ii*), and errors in the estimation of number of atoms can lead to an error of ~30% in the determination of $\beta_{Cs,Li}$.

If the MOTs are loaded to a steady state with $N_{A,0}$ and $N_{B,0}$ number of atoms and then the loading of $A$ suddenly stopped by setting $L_A = 0$, Eq. (1) takes the following form when assumptions (*i*) and (*ii*) are valid:

$$\frac{dN_A}{dt} = -(\gamma_A + \beta_{A,B} n_{B0}) N_A - \frac{\beta_A}{\sqrt{8} V_A} N_A^2 \quad [2]$$

The solution of Eq. (2) is:

$$N_A(t) = \frac{N_{A0} e^{-\kappa t}}{1 + \frac{N_{A0}}{\sqrt{8} V_A} \frac{\beta_A}{\kappa} (1 - e^{-\kappa t})} \quad [3]$$

where $\kappa = \gamma_A + \beta_{A,B} n_{B0}$. Note that in the absence of the MOT of species $B$, the value of $\kappa$ yields the value of $\gamma_A$. The fits of the experimental data to Eq. (3) are shown in Fig. 7.

The values of $\gamma_{Cs}$ and $\beta_{Cs}$ are determined by fitting the decay of the $^{133}$Cs atom in absence of the $^{7}$Li MOT to Eq. (3) with $n_{Li0}$ set to zero. The value of $\kappa$ (= $\gamma_{Cs} + \beta_{Cs,Li} n_{Li0}$) is determined by fitting the decay of the $^{133}$Cs atom in presence of the $^{7}$Li MOT to Eq. (3). The value of $\beta_{Cs,Li} = (\kappa - \gamma_{Cs})/n_{Li0}$ is then determined using the measured values of $\kappa$, $\gamma_{Cs}$ and $n_{Li0}$. At our standard operating conditions discussed in section II, we find $\beta_{Cs}$ to be ~ 2.5(6)×10$^{-11}$ cm$^3$s$^{-1}$ and $\beta_{Cs,Li}$ to be ~ 7.8(8)×10$^{-12}$ cm$^3$s$^{-1}$. Notably, we also measured the decay of $^{7}$Li atoms in absence a $^{133}$Cs MOT (see Fig. 8) and obtained $\beta_{Li}$ to be ~ 1.2(2)×10$^{-12}$ cm$^3$s$^{-1}$ following a similar fitting method. We did not measure $\beta_{Li,Cs}$ but that can also be done by changing the operating parameters to make the $^{7}$Li atomic cloud much smaller than the $^{133}$Cs cloud.

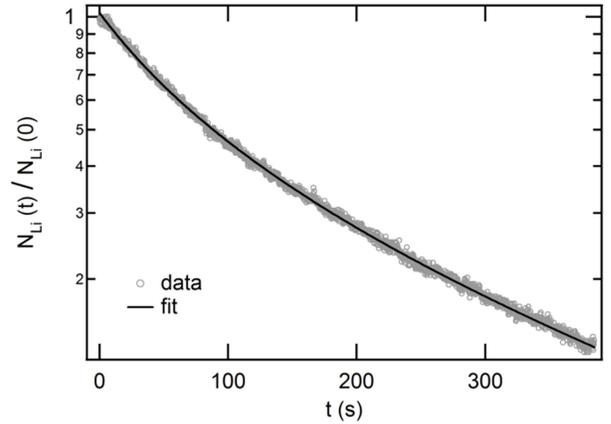

**FIG. 8.** The decay in the number of $^{7}$Li atoms when the loading of atoms is stopped by closing the atomic beam shutter. The solid line is a fit of the data to Eq. (3).

The loss of atoms due to collisions in a MOT can arise from a variety of inelastic collision processes such as hyperfine-changing collisions (HCC), fine-structure changing collisions (FCC) and radiative escape (RE) [58–60]. Binary collisions between atoms in the electronic ground state can result in HCC. However, the trap depth of our MOTs are large compared to the ground state hyperfine splitting of $^{133}$Cs and $^7$Li, $h \times 9.2$ GHz and $h \times 0.8$ GHz, respectively, and thus the HCC cannot lead to atom loss. Binary collisions involving both atoms in the electronic excited state are rare because of the low density of excited-state atoms and the short lifetime of atoms in the excited state. Thus, the primary contributor to loss of atoms in a MOT is binary collisions involving one atom in the electronic ground state and the other in an electronic excited state e.g. FCC and RE. Individual contributions of FCC and RE are difficult to ascertain but both depend on collisions between an excited state atom and a ground state atom. Li atoms in the excited $2p\ ^2P_{3/2}$ state (denoted by Li*) and Cs atoms in the $6p\ ^2P_{3/2}$ state (denoted by Cs*) exist in the dual species MOT. The Li*-Cs interaction potential is repulsive at long-range and therefore it prevents Li* and Cs from coming close enough (~10 Å) for FCC or RE to occur [58]. The Li-Cs* interaction potential is attractive at long-range and can allow FCC or RE to occur. To test if this is indeed the case, we perform experiments at different detuning $\delta_{Cs}$ of the Cs cooling laser beams. As shown in Fig. 9(a), we find that $\beta_{Cs,Li}$ decreases from $1.2 \times 10^{-11}$ cm$^3$s$^{-1}$ to $7.1 \times 10^{-12}$ cm$^3$s$^{-1}$ on changing $\delta_{Cs}$ from -6 MHz to -12 MHz. The correlation between $\delta_{Cs}$ and loss rate coefficient supports FCC and RE as the loss mechanism in the case of $^7$Li-$^{133}$Cs collisions, as discussed below. This implies that collision induced losses can be supressed by using a dark spot MOT for $^{133}$Cs so that there is negligible excitation to the $6p\ ^2P_{3/2}$ state. Such a strategy has previously been implemented in a $^7$Li-$^{85}$Rb dual species MOT [61]. In sharp contrast to $\beta_{Cs,Li}$, the value of $\beta_{Cs}$ remains almost unchanged when $\delta_{Cs}$ changes from -6 MHz to -12 MHz as shown in Fig. 9(b).

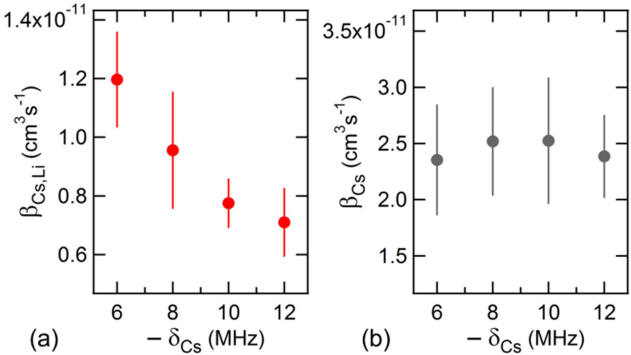

**FIG. 9.** The values of collision rate coefficients $\beta_{Cs,Li}$ (panel a) and $\beta_{Cs}$ (panel b) measured at different values of the Cs cooling laser detuning $\delta_{Cs}$.

The difference in the trends of $\beta_{Cs}$ and $\beta_{Cs,Li}$ is attributed to the long-range interaction potentials ($U_L$) being different for homo-nuclear Cs-Cs* ($U_L = -C_3/r^3$) and hetero-nuclear Li-Cs* ($U_L = -C_6/r^6$) cases. During a two-body collision, the interaction shifts the energy levels of the atoms and therefore, for a fixed value of $\delta_{Cs}$, the probability of excitation to a Cs-Cs* or Li-Cs* collision-complex is maximized at a particular inter-nuclear separation $r_C$ called the Condon point where the detuning compensates for the shift in energy level. The typical value of $r_C$ is of the order $10^3$ Å for the $-C_3/r^3$ potential (Cs-Cs* case) and of the order $10^2$ Å for the $-C_6/r^6$ potential (Li-Cs* case), while FCC and RE occur in the short-range part of the potential requiring inter-nuclear separation of the order 10 Å. This implies that the inter-nuclear separation of the long-range Cs-Cs* or Li-Cs* collision-complex must decrease to ~ 10 Å within the lifetime of the complex (lifetime ~ 30 ns, similar to that of Cs* atom). Noting that the typical speed of Cs atoms in a MOT is ~1 m/s, an excited complex can travel ~ 300 Å before it decays. In the case of Cs-Cs* collisions, as $\delta_{Cs}$ changes from -6 MHz to -12 MHz, the value of $r_C$ reduces which increases the probability that the excited complex samples the short-range part of the potential. However, the change in $r_C$ is small because the $-C_3/r^3$ potential varies relatively slowly with $r$ compared to the $-C_6/r^6$ potential. The unchanged value of $\beta_{Cs}$ with varying $\delta_{Cs}$ is thus explained by the relatively unchanged value of $r_C$. On the other hand, in the case of Li-Cs* collisions, the value of $r_C$ is already so small that the excited complex can always sample the short-range part of the potential irrespective of the detuning. This would suggest that $\beta_{Cs,Li}$ should be independent of $\delta_{Cs}$. However, the probability that the cooling light excites an isolated Cs atom to Cs* decreases as $\delta_{Cs}$ changes from -6 MHz to -12 MHz. Therefore, the probability of Li-Cs* collisions decrease as seen in the experiment.

We note that our value of $\beta_{Cs}$ is slightly higher than the previously reported value (~$1.5 \times 10^{-11}$ cm$^3$s$^{-1}$) in Ref. [59] possibly because the intensity of the cooling laser in our experiments is ~10 times higher, thus increasing the fraction of Cs* atoms which are primarily responsible for collision-induced losses as discussed above. Our value of $\beta_{Cs,Li}$ is a factor of ~7 lower than the previously reported value in Ref. [60] possibly because the trap depth of our Cs MOT is higher (owing to the higher light intensity). Our value of $\beta_{Li}$ is in excellent agreement with earlier works carried out under similar conditions [55,56].

### B. Collisions between ions and neutral atoms

We study collisions between $^{133}$Cs$^+$ and $^{133}$Cs and, separately, between $^7$Li$^+$ and $^7$Li. The measurement technique in both cases is the same and is based on the methodologies developed in Refs. [62–64]. To determine the Cs-Cs$^+$ (Li-Li$^+$) collision rate, three experiments are performed. In the first experiment, a Cs (Li) MOT is loaded to a steady state in the absence of photo-ionization as shown with grey open circles in Fig. 10(a) (Fig. 10(b)). In the second experiment, a Cs (Li) MOT is loaded to a steady state in presence of a 505-nm (266-nm) light source that causes photo-ionization of Cs (Li) atoms from the $6p\ ^2P_{3/2}$ ($2p\ ^2P_{3/2}$) state. The corresponding Cs (Li) MOT loading data is shown with blue filled circles in Fig. 10(a) (Fig. 10(b)). In the third experiment, a Cs (Li) MOT is loaded to a steady state while the photo-ionization light source and the ion trap voltages are both kept on. The corresponding Cs (Li) MOT loading data is shown with red squares in Fig. 10(a) (Fig. 10(b)). The data is interpreted as follows. The photo-ionization of atoms causes loss of atoms from the MOT at a rate $\xi$. A fraction of the ions formed by photo-ionization are trapped in the Paul trap. The competition between the continuous loading of ion by photo-ionization and the loss of ions due to ion-ion rf heating leads to a steady state with a fixed number of ions in the Paul trap. The trapped ions collide with the atoms leading to atom loss, resulting in a steady state MOT with a reduced number of atoms. The additional loss of Cs (Li) atoms due the presence of trapped $Cs^+$ ($Li^+$) is used to determine the $Cs-Cs^+$ ($Li-Li^+$) collision rate following the same rate equation model described by Eq. (1). In this case, the species $A$ is identified as Cs (Li) and species $B$ is identified as $Cs^+$ ($Li^+$).

To simply the analysis, we again make two assumptions: (*i*) The Cs (Li) atomic cloud is much smaller in size and completely immersed in cloud of trapped $Cs^+$ ($Li^+$) whose density $n_{Cs+0}$ ($n_{Li+0}$) remains fixed. This allows $\beta_{A,B} \int n_A n_B d^3r$ to be written as $\beta_{A,B} n_{B0} N_A$. The ion clouds in our Paul traps are typically ellipsoidal with both major and minor axis exceeding 1.5 mm, as determined from SIMION simulations. This implies that the ion cloud is much larger than the atomic cloud. Additionally, to ensure that the density of trapped ions does not change while performing the third experiment, we first saturated the ion trap (by keeping the MOT, photo-ionization source and the ion trap on for a prolonged time) and then briefly switch off the MOT laser beams for 50-100 ms to empty the MOT, followed by switching on the laser beams to record the MOT loading dynamics. This ensures that ion trap is constantly being replenished, except during the 50-100 ms duration, and thus the density of trapped ions remains approximately fixed. It is implicitly assumed that each atom-ion collision leads to the loss of the atom from the MOT because the MOT is shallow. (*ii*) We assume that we can write $\beta_A \int n_A^2 d^3r = \beta_A N_A^2/\sqrt{8}V_A \approx \beta_A n_{A0} N_A$. Note that $\beta_{Cs}$ ($\beta_{Li}$) values were obtained in section III.A. In writing the final simplifying step we implicitly assumed the MOT density to be constant which is only approximately valid in the experiment and therefore may lead to some errors in the analysis. However, these errors are small as will be clear from the good quality of the fits shown below. Had we not made the last approximation, the differential equation would be non-linear with a term dependent on $N_A^2$ and the analytical solution would be a complicated expression (see Appendix) that might increase the possibility of the fitted parameters being correlated.

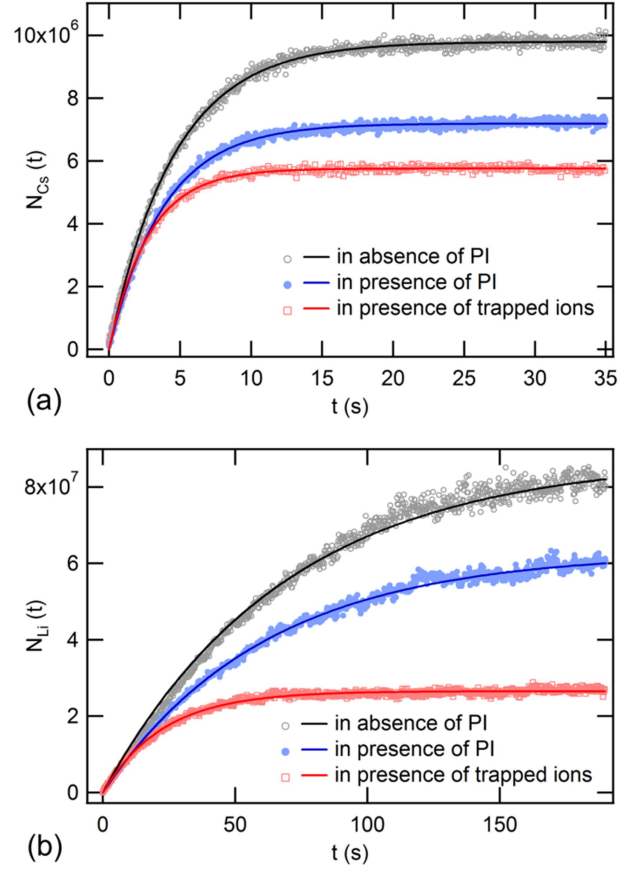

**FIG. 10.** The loading of the $^{133}$Cs MOT (panel a) and $^7$Li MOT (panel b) under three different conditions are shown. The steady state atom number in a MOT is lower in presence of photo-ionization (PI) and is further lowered in presence of trapped ions. The solid lines are fits to the data as described in text.

On making the two assumptions, the rate equation simplifies to:

$$\frac{dN_A}{dt} = L_A - (\gamma_A + \xi_A + \beta_A n_{A0} + \beta_{A,B}\ n_{B0})\ N_A \quad [4]$$

Here, we identify species $A$ with Cs (Li) and species $B$ with $Cs^+$ ($Li^+$). $\gamma_A$ and $\xi_A$ represent the atom loss rate due collisions with background gas atoms and photo-ionization, respectively. $\beta_{A,B}$ represents the loss rate coefficient that accounts for loss of atoms due to collisions with ions and this last term contributes only when the ions are trapped. For brevity we define $\kappa_A = \gamma_A + \xi_A + \beta_A n_{A0}$ and $\gamma_{A,B} = \beta_{A,B}\ n_{B0}$. Here $\gamma_{A,B}$ is the atom-ion collision rate.

In the first experiment, in absence of photo-ionization $\xi_A = 0$ and $n_{B0} = 0$, the solution to Eq. (4) is $N_A(t) = \frac{L_A}{\gamma_A + \beta_A n_{A0}}(1 - e^{-(\gamma_A + \beta_A n_{A0})\ t})$. From fits to the data shown with black lines in Fig. 10(a) and Fig. 10(b), we determine $L_{Cs} \sim 2.2 \times 10^6$ s$^{-1}$ and $\gamma_{Cs} + \beta_{Cs} n_{Cs0} \sim 0.22$ s$^{-1}$ for Cs-Cs$^+$ case and $L_{Li} \sim 1.3 \times 10^6$ s$^{-1}$ and $\gamma_{Li} + \beta_{Li} n_{Li0} \sim 0.014$ s$^{-1}$ for Li-Li$^+$ case. In the second experiment, where photo-ionization is present but the ion trap is switched off, only the last term in Eq. (4) is dropped and the solution is $N_A(t) = \frac{L_A}{\kappa_A}(1 - e^{-\kappa_A t})$. From fits to the data shown with blue lines in Fig. 10(a) and Fig. 10(b), we determine $\kappa_{Cs} \sim 0.26$ s$^{-1}$ for the Cs-Cs$^+$ case and $\kappa_{Li} \sim 0.016$ s$^{-1}$ for the Li-Li$^+$ case, respectively. In the third experiment, where the ion trap is kept on, all terms in Eq. (4) contribute and the solution is $N_A(t) = \frac{L_A}{\kappa_A + \gamma_{A,B}}(1 - e^{-(\kappa_A + \gamma_{A,B}) t})$. From fits to the data shown with red lines in Fig. 10(a) and Fig. 10(b), we determine $\kappa_{Cs} + \gamma_{Cs,Cs+} \sim 0.36$ s$^{-1}$ for the Cs-Cs$^+$ case and $\kappa_{Li} + \gamma_{Li,Li+} \sim 0.044$ s$^{-1}$ for the Li-Li$^+$ case. With the values of $\kappa_{Cs}$ and $\kappa_{Li}$ previously determined, we find $\gamma_{Cs,Cs+} = 0.10(2)$ s$^{-1}$ and $\gamma_{Li,Li+} = 0.028(5)$ s$^{-1}$ for Cs-Cs$^+$ and Li-Li$^+$ collisions, respectively. It is noted that $\gamma_{Cs,Cs+}$ is higher than $\gamma_{Li,Li+}$ because Cs is more polarizable than Li [65] which increases the atom-ion interaction (i.e. the $C_4$ coefficient) in the Cs-Cs$^+$ case leading to a higher collision rate.

To determine the atom-ion collision rate coefficient $\beta_{A,B}(= \gamma_{A,B}/n_{B0})$, the density of trapped ions ($n_{B0}$) needs to be estimated. It is difficult to measure $n_{B0}$ experimentally because the ions are optically inactive and cannot be imaged using fluorescence detection. The alternative approach is to experimentally estimate the number of trapped ions ($N_{B0}$) and divide it by the volume ($V_B$) of the trapped ions as determined from simulations. We estimate $V_{Cs+} \sim 0.2$ cm$^3$ and $V_{Li+} \sim 0.7$ cm$^3$ from SIMION simulations. Unfortunately, we found it difficult to accurately estimate the number of ions by detecting using the axial CEM because the ion cloud is elongated, restricting all ions from being detected equally efficiently. Determining the ion density in our experiment remains an important future plan. However, on using a typical value of ion density in such traps i.e. $n_{B0} \sim 2\times10^6$ cm$^{-3}$ [62–64], we get $\beta_{Cs,Cs+} \sim 0.5\times10^{-7}$ cm$^3$s$^{-1}$ and $\beta_{Li,Li+} \sim 1.4\times10^{-8}$ cm$^3$s$^{-1}$. These value of $\beta_{Cs,Cs+}$ is of the same order of magnitude as other alkali-metal species [62–64]. The value of $\beta_{Li,Li+}$ has not been measured earlier.

## C. Sympathetic cooling of ions by atoms

The ions trapped in a Paul trap rapidly gain energy due to ion-ion rf heating [66] and are eventually lost from the trap when the ion trajectory expands beyond the trappable region. Sympathetic cooling of the ions by collisions with cold atoms counteracts the gain in ion energy and arrests the spatial expansion of the ion leading to enhanced lifetime of the ion in the Paul trap. Increase in the lifetime of the ion in a Paul trap is therefore a signature of ion cooling [22,24,32,67]. In homonuclear atom-ion collisions such as $^{133}$Cs-$^{133}$Cs$^+$ and $^7$Li-$^7$Li$^+$ collisions, there are two mechanisms that can cool the ions, namely, elastic collisions and resonant charge exchange. In cooling by elastic collisions [22], a fast ion loses a fraction of its kinetic energy during a collision with a cold atom essentially at rest. The fraction of energy lost per collision depends on the mass ratio between the atom and the ion – repeated elastic collisions lead to substantial cooling of the ion. Resonant charge exchange occurs only when the ion $A^+$ and atom $A$ are of the same species, in which case an electron can transfer seamlessly from the atom to the ion without costing any energy. If resonant charge exchange occurs between a fast ion $A^+$ and a cold atom $A$, an ion $A^+$ at rest is produced which remains trapped in the Paul trap while the newly produced fast atom $A$ escapes from the MOT. Since a single collision is sufficient to produce an ion at rest, the cooling by resonant charge exchange is an extremely efficient process [24]. Earlier works have demonstrated cooling of ions to ~10 meV by collisions with laser cooled atoms [24].

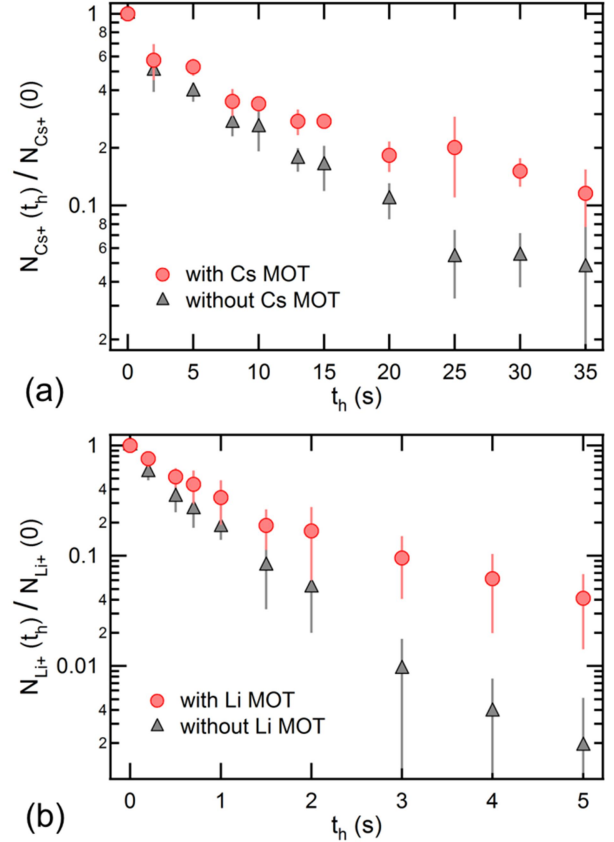

**FIG. 11.** The decay of $^{133}$Cs$^+$ (panel a) and $^7$Li$^+$ (panel b) from the ion trap in absence (triangles) and presence (circles) of atoms in a MOT are shown. The increase in lifetime of ions in the presence of atoms in a MOT is evident.

To study the collisional cooling of Cs$^+$ (Li$^+$), we compare the lifetime of Cs$^+$ (Li$^+$) in the ion trap either in absence or presence of the MOT of Cs (Li). The experimental sequence is the following. The Cs (Li) MOT is loaded to steady state and the ion trap voltages are turned on. The ion trap is then loaded with Cs$^+$ (Li$^+$) by turning on the photo-ionization light. The instant at which the photo-ionization light switches off is defined to be the zero of the time sequence. The Cs$^+$ (Li$^+$) ions are then held in the ion trap for a predetermined hold time ($t_h$), either in absence or presence of the Cs (Li) atoms, and then extracted from the ion trap by switching off the dc voltage on end cap #2. The number ($N(t_h)$) of ions hitting the axial CEM, which is proportional to the number of ions that survived in the ion trap, is counted. The ratio $N(t_h)/N(0)$ is plotted against $t_h$ for the Cs-Cs$^+$ and Li-Li$^+$ cases in Fig. 11(a) and Fig. 11(b), respectively. In both cases it is seen that the lifetime of the ions increases when the ions are held in presence of cold atoms which is a tell-tale signature of sympathetic cooling of ions by collisions with atoms [67]. It is also seen that the enhancement of lifetime is more pronounced in the Cs-Cs$^+$ case compared to the Li-Li$^+$ case. We conjecture that collisions of Li$^+$ with the trace amounts of un-trapped background atoms of Cs and other species which are heavier than Li$^+$ leads to an additional heating channel for Li$^+$ since the atom/ion mass ratio is higher than the critical mass ratio (= 1) for ion cooling by a uniform buffer gas [68]. The additional heating limits the lifetime of Li$^+$ in the trap.

The data in Fig. 11(a) and Fig. 11(b) cannot differentiate between the role of elastic collision and resonant charge exchange collisions in the sympathetic cooling of ions by atoms. Differentiating between these two mechanisms will be the topic of a future experiment.

## IV. DISCUSSIONS

We describe the construction and operation of a versatile apparatus that is capable of simultaneously trapping several species of atoms and ions. We discuss innovations in the design that simplify technical requirements and enable robust operation of the system. The dual-species oven operates at lower temperatures compared to earlier reports on Li and can be adapted for other species. The use of a short Zeeman slower and operation of a Li MOT with a single laser are useful for compact setups. We show operation of 2D and 3D MOTs of Cs using common laser sources. We show trapping of ions in a relatively big ion trap with high optical access which is well-suited for studies of low energy atom-ion collisions. The capability to detect ions using a ToF MS is also demonstrated.

We report studies of low energy atom-atom and atom-ion collisions in the apparatus. We demonstrate the versatility of the apparatus by studying $^7$Li-$^{133}$Cs, $^{133}$Cs-$^{133}$Cs$^+$ and $^7$Li-$^7$Li$^+$ collisions. We observe signatures of sympathetic cooling of $^7$Li$^+$ by $^7$Li and $^{133}$Cs$^+$ by $^{133}$Cs. The $^7$Li-$^7$Li$^+$ system is the lightest atom-ion system among alkali species and is an important system for studies of charge transport in the quantum regime because the *s*-wave regime in atom-ion collisions is attainable at a relatively higher temperature (~ 24 μK). In the future, the apparatus can be used to produce ultracold LiCs molecules by associating Li and Cs, which will enable studies of low energy collisions between ions and polar LiCs molecules. It may be noted that the LiCs molecule has the highest dipole moment among all bi-alkali molecules and may enable further studies of long-range interacting quantum systems. It will also be possible to study collisions among LiCs molecules and use REMPI to detect the formation of long-lived collisional complexes which have been conjectured to limit the trap lifetime of other cold molecules [39,69–72].

## Acknowledgements

We thank the TIFR Central Workshop for building several components of the apparatus. We acknowledge the contributions of Kamal Kumar, Vishu Gupta, Sagar Dam, Gorachand Das and Manojit Das as project students during different stages of this work. We acknowledge funding from the Department of Atomic Energy, Government of India, under Project Identification No. RTI4002.

## Author Contributions

S.D. conceived the idea and designed the apparatus and the experiments. S.D. and B.R. built the apparatus. B.R. performed the simulations with inputs from S.D.. B.R. and S.B. performed the experiments. B.R. analyzed the data on atom-atom and atom-ion collisions. S.B. analyzed the data on sympathetic cooling. All authors discussed the analysis and the results. S.D. wrote the manuscript with inputs from B.R..

## Data availability

The data that support the findings of this study are available from the corresponding authors upon reasonable request.

## Appendix

We briefly discuss the analytical solution in the case where $\beta_A \int n_A^2 d^3r = \beta_A N_A^2/\sqrt{8}V_A$ and the approximation $\approx \beta_A n_{A0} N_A$ is not made. The rate equation for atom-ion collisions discussed in section III.B is then modelled as:

$$\frac{dN_A}{dt} = L_A - (\gamma_A + \xi_A + \beta_{A,B}\, n_{B0}) N_A - \frac{\beta_A}{\sqrt{8}V_A} N_A^2 \quad [A1]$$

For brevity we define $\tilde{\kappa}_A = \gamma_A + \xi_A + \beta_{A,B}\, n_{B0}$ and $\tilde{\beta}_A = \beta_A/\sqrt{8}V_A$. The solution to Eq. (A1), considering $N_A(0) = 0$, is:

$$N_A(t) = -\frac{\tilde{\kappa}_A}{2\tilde{\beta}_A} + \frac{\sqrt{4L_A\tilde{\beta}_A + \tilde{\kappa}_A^2}}{2\tilde{\beta}_A} \times$$

$$tanh\left(\sqrt{4L_A\tilde{\beta}_A + \tilde{\kappa}_A^2}\,\frac{t}{2} + tanh^{-1}\left(\frac{\tilde{\kappa}_A}{\sqrt{4L_A\tilde{\beta}_A + \tilde{\kappa}_A^2}}\right)\right) \quad [A2]$$

The expression in Eq. (A2) describes the loading of the MOT in presence of trapped ions of density $n_{B,0}$. On setting $n_{B,0} = 0$, i.e. $\tilde{\kappa}_A \rightarrow \gamma_A + \xi_A$, the same expression describes the loading of the MOT in the absence of trapped ions and the MOT loading data in absence of trapped ions can be fit to extract $\gamma_A + \xi_A$ and $\tilde{\beta}_A$. In practice, we found orders of magnitude variations in the fitted value of $\tilde{\beta}_A$ because it is many orders of magnitude smaller than $\gamma_A + \xi_A$, making the fit unreliable. Therefore, we constrain $\tilde{\beta}_A$ using the value of $\beta_A$ obtained in section III.A. With $\gamma_A + \xi_A$ and $\tilde{\beta}_A$ held fixed, the MOT loading data in the presence of trapped ions can be fit to Eq. (A2) to extract $\tilde{\kappa}_A = \gamma_A + \xi_A + \beta_{A,B} \, n_{B0}$. Thus the atom-ion collision rate $\gamma_{A,B} = \beta_{A,B} \, n_{B0}$ can be determined. On using this method, we get $\gamma_{Cs,Cs+} \sim 0.13$ s$^{-1}$ and $\gamma_{Li,Li+} \sim 0.031$ s$^{-1}$ which are in reasonable agreement with the values obtained in section III.B.